\documentclass[aps, prb, twocolumn,notitlepage, amsmath, amssymb, superscriptaddress, nofootinbib]{revtex4-1}

\makeatletter
\def\p@subsection{}
\def\p@subsubsection{}
\makeatother

\usepackage[utf8]{inputenc}
\usepackage{graphicx}
\usepackage{units}
\usepackage{color}
\usepackage[pdftex, colorlinks=true, linkcolor=myblue, citecolor=myblue,
urlcolor=myblue]{hyperref}
\usepackage{times}
\usepackage{comment}
\usepackage{graphicx}
\usepackage{amsmath, calc}
\usepackage{mathrsfs}
\usepackage{MnSymbol}
\usepackage{amsfonts}
\usepackage{amssymb}
\usepackage{bm}
\usepackage{bbm}
\usepackage{color}
\usepackage{array}
\usepackage{units}
\usepackage{mathrsfs,amsmath}

\definecolor{myblue}{rgb}{0,0,1}
\definecolor{myred}{rgb}{1,0,0}

\newcommand{\expect}[1]{\langle #1 \rangle}

\DeclareMathOperator{\Tr}{Tr}

\begin{document}
\title{Chiral current circulation and $\mathcal{PT}$ symmetry in a trimer of oscillators}

\author{Charles A. Downing}
\affiliation{Instituto de Nanociencia y Materiales de Arag\'{o}n (INMA), CSIC-Universidad de Zaragoza, Zaragoza 50009, Spain}
\affiliation{Departamento de F\'{i}sica de la Materia Condensada, Universidad de Zaragoza, Zaragoza 50009, Spain}
\affiliation{Department of Physics and Astronomy, University of Exeter, Exeter EX4 4QL, United Kingdom}

\author{David Zueco}
\affiliation{Instituto de Nanociencia y Materiales de Arag\'{o}n (INMA), CSIC-Universidad de Zaragoza, Zaragoza 50009, Spain}
\affiliation{Departamento de F\'{i}sica de la Materia Condensada, Universidad de Zaragoza, Zaragoza 50009, Spain}

\author{Luis Mart\'{i}n-Moreno}
\affiliation{Instituto de Nanociencia y Materiales de Arag\'{o}n (INMA), CSIC-Universidad de Zaragoza, Zaragoza 50009, Spain}
\affiliation{Departamento de F\'{i}sica de la Materia Condensada, Universidad de Zaragoza, Zaragoza 50009, Spain}

\date{\today}

\begin{abstract}
We present a simple quantum theory of a bosonic trimer in a triangular configuration, subject to gain and loss in an open quantum systems approach. Importantly, the coupling constants between each oscillator are augmented by complex arguments, which give rise to various asymmetries. In particular, one may tune the complex phases to induce chiral currents, including the special case of completely unidirectional (or one-way) circulation when certain conditions are met regarding the coherent and incoherent couplings. When our general theory is recast into a specific non-Hermitian Hamiltonian, we find interesting features in the trimer population dynamics close to the exceptional points between phases of broken and unbroken $\mathcal{PT}$ symmetry. Our theoretical work provides perspectives for the experimental realization of chiral transport at the nanoscale in a variety of accessible nanophotonic and nanoplasmonic systems, and paves the way for the potential actualization of nonreciprocal devices.
\end{abstract}


\maketitle


\section*{Introduction}
\label{Intro}

The typical custom of greeting friends and colleagues with a same-handed handshake is possible because of the intrinsic chirality of the human hand. Since left and right hands are non-superimposable mirror images of one another, different-handed handshakes are thankfully rare. Handedness (or chirality) in condensed matter physics is linked to a multitude of rather more interesting phenomena. Recent examples include the chirality of electrons in graphene, which leads to extraordinary transport properties~\cite{Katsnelson2006}, the chiral edge states in photonic crystals, which support a version of the quantum Hall effect~\cite{Raghu2008}, and the chiral anomaly associated with Weyl fermions, which engenders topological Fermi arcs~\cite{Xu2015}.

When investigating chirality and its interplay with light at the nanoscale, the toolbox of chiral quantum optics is of great utility~\cite{Lodahl2017, Andrews2018}. The field seeks to describe quantum optical systems where the interactions are asymmetric, and in extreme cases completely unidirectional (or one-way)~\cite{Gardiner2004}. The extra freedom gained via the exploitation of chiral coupling can be employed in a new generation of nanophotonic devices~\cite{Chang2018, Amico2019, Huang2020, Carusotto2020, Bekenstein2020}, for example in chiral waveguides~\cite{Roy2017, Bello2019, Sanchez2019}, which promise unprecedented control of the flow and transmission of excitations at the smallest scales. 

Recently, a groundbreaking experiment led by a team at Google reported the realization of a synthetic magnetic field in a trimer of superconducting qubits~\cite{Roushan2017}. The required phases associated with the qubit-qubit interactions arose by sinusoidally modulating the couplings, mimicking the effects of the Peierls phases which appear due to the application of a real magnetic field~\cite{Harper1955, Hofstadter1976}. It was shown how chiral ground-state currents can be engineered, which has significant implications for potential nonreciprocal nanophotonic devices, such as optical isolators~\cite{Jalas2013, Sollner2015, Sayrin2015, Zhang2018} and circulators~\cite{Scheucher2016, Barzanjeh2017, Shen2018, Ruesink2018}. The trailblazing experiment of Ref.~\cite{Roushan2017} directly inspired the theoretical work presented here.

In our quantum theory, we consider a triangle of harmonic oscillators in an open quantum systems approach. The oscillators are primarily coupled through coherent coupling, which in general is described by a complex-valued parameter in order to capture the effects of an accumulated phase in the triangular loop. Dissipation and gain are introduced incoherently via a quantum master equation~\cite{Jin2013, Quijandria2018, Jaramillo2020}, from which we derive the population dynamics and current in the system~\cite{Gardiner2014}.

As the first application of our theory, we consider the specific situation where our general model maps onto a $\mathcal{PT}$ symmetric system, with balanced loss and gain~\cite{Bender2018, Quijandria2018, Konotop2016, Ashida2020, Bergholtz2020}. The introduction of $\mathcal{PT}$ symmetric quantum mechanics into photonics has led to insightful studies of exceptional points and symmetry breakdowns, which are typically associated with nontrivial phenomena~\cite{Regensburger2012, Ganainy2018, Ozdemir2019, Miri2019, Chen2020}. While prior studies of $\mathcal{PT}$ symmetric trimers have primarily focused on non-Hermitian Hamiltonian approaches~\cite{Li2011, Duanmu2013, Li2013, Suchkov2016, Suchkov2016b, Xue2017, Leykam2017, Du2018, Jin2019}, here we upgrade the analysis to a full master equation. We find the accumulated hopping phase in our system crucially determines the existence of broken and unbroken $\mathcal{PT}$ symmetric phases, and we reveal a universal ratio of relative coherent-coupling-to-loss strength above which $\mathcal{PT}$ symmetric phases are possible (for at least some subset of accumulated phase). Our analysis has important consequences for population dynamics in looped systems, which can display both convergent and divergent behaviors.

The second implementation of our theory relaxes the restraints required for the $\mathcal{PT}$ symmetric setup of the trimer. We investigate in detail how modulating both the accumulated phase and coherent coupling strengths effects the population transfer in the triangle. In particular, we find instances of chiral currents as soon as the accumulated phase is nonzero, and we reveal especially strong features when the phase is equal to $\pi/2$. 

\begin{figure*}[tb]
 \includegraphics[width=0.95\linewidth]{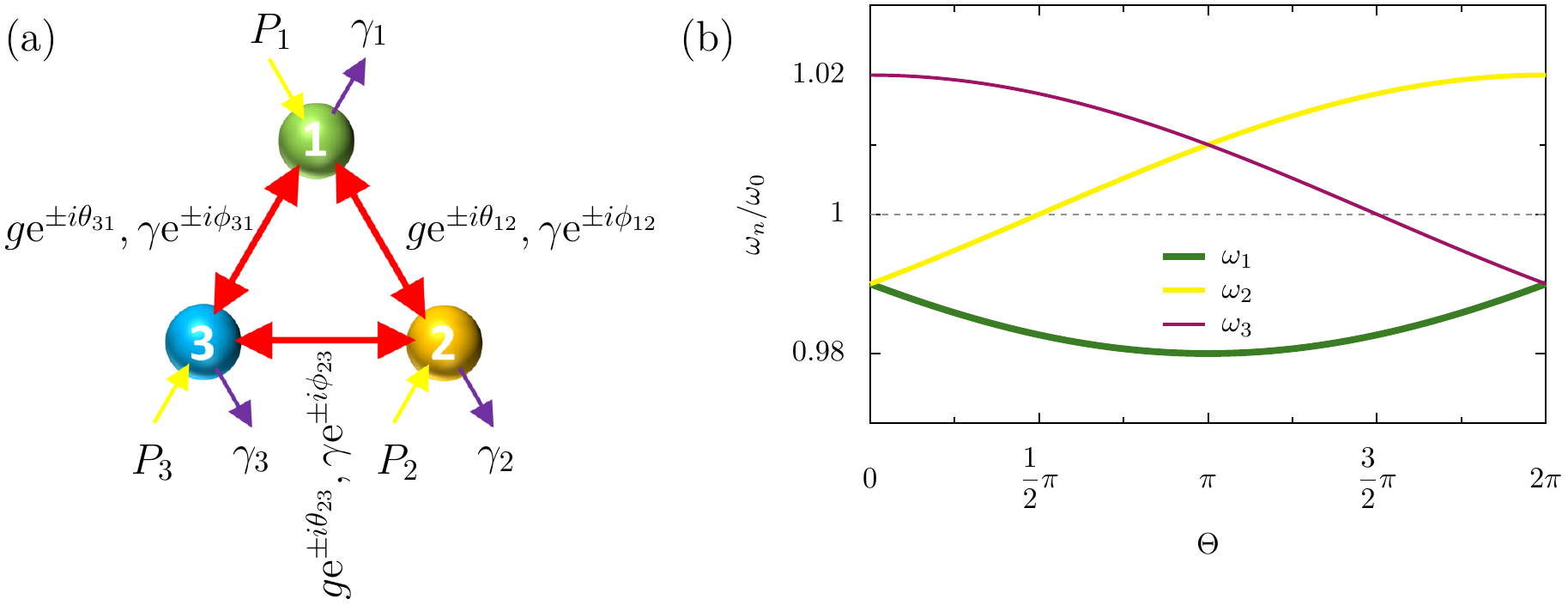}
 \caption{ Panel (a): a sketch of the triangular arrangement of oscillators, where each oscillator is of resonance frequency $\omega_0$ and the $n$-th oscillator is subject to gain $P_n$ (yellow arrows) and loss $\gamma_n$ (purple arrows). The magnitude of the three coherent (dissipative) coupling constants is $g$ ($\gamma$), which are augmented with complex arguments $\theta_{nm}$ ($\phi_{nm}$), as is labelled near to the red arrows. Panel (b): the eigenfrequencies $\omega_n$ of the trimer system, as a function of the cumulative phase $\Theta$, where $g = \omega_0/100$ [cf. Eq.~\eqref{eq:Eig_one}]. The gray dashed line is a guide for the eye at $\omega_0$. }
 \label{sketch}
\end{figure*}

The third and final application of our theory is related to cascaded quantum systems, as independently pioneered by Gardiner and Carmichael in the early 1990's~\cite{Gardiner1993, Carmichael1993}. Cascaded systems are those where the output field from one body drives another body, strictly without any backaction. In order to achieve this unidirectional coupling in our trimer of oscillators, we need to introduce another quantity: incoherent coupling~\cite{Akram2000, Wang2020}. The careful balancing of both the relative strengths and relative phases of the coherent and dissipative couplings in the triangle allows us to map our quantum master equation onto the celebrated cascaded quantum system equations~\cite{Gardiner1993, Carmichael1993}. Therefore, we unveil perfectly unidirectional circulation in the trimer of oscillators.

Our theory complements a growing body of work on the interplay of non-Hermitian systems and synthetic magnetic fields, with recent works showing unidirectional transport in long chains of resonators~\cite{Jin2016, Jin2018, Jin2019b, Wang2019b, Zhang2019, Zhang2020}. We would also like to mention interesting studies of four coupled resonators~\cite{Jin2016b, Jin2018b}, a system which has recently been ingeniously probed experimentally using acoustic cavities~\cite{Ding2016}.

The synthetic gauge fields which are required to give rise to phase-dependent hopping parameters may be realized in a variety of ways. In their study of coupled cavity arrays, Umucalilar and Carusotto showed how non-trivial phase associated with tunneling photons can be induced through the introduction of optically active materials~\cite{Umucalilar2011}. The photonic analogue of the Aharonov-Bohm effect was described in Refs.~\cite{Fang2012, Fang2012b, Fang2013}, which exploited the phase of modulation of a dynamically changing photonic system in order to generate effective gauge potentials. Furthermore, the pioneering experiment of Tzuang and co-workers used an on-chip Ramsey-type interferometer to create an effective magnetic field for photons~\cite{Tzuang2014}. It was also proposed theoretically that synthetic magnetic fields can be realized in lattices of resonators without modulating coupling rates, by exploiting transverse index gradients~\cite{Longhi2013}. Phenomena such as Bose-Einstein condensation~\cite{Kennedy2015}, Landau levels~\cite{Schine2016} and chiral magnetism~\cite{Wang2019} have already been observed in synthetic magnetic fields, highlighting the remarkable capabilities of this active research field.


\section*{Model}
\label{Model}

Our minimal model of chiral currents at the nanoscale is based upon a triangle of harmonic oscillators, as sketched in Fig.~\ref{sketch}~(a), which has  applications across a plethora of circuit QED, ultracold atom, nanophotonic and plasmonic platforms. We utilize an open quantum systems approach to allow us to conveniently describe loss and gain in the system. This framework leads to the Hamiltonian dynamics described in what follows, and a corresponding quantum master equation, which is introduced afterwards. We then calculate the mean population dynamics in the trimer, and the corresponding local and global currents.


\subsection*{Hamiltonian}
\label{Hamm}

The Hamiltonian operator for the system reads ($\hbar = 1$ throughout)
\begin{align}
\label{eq:Ham}
\hat{H} =& ~\omega_0 \left( b_{1}^{\dagger} b_{1}  + b_{2}^{\dagger} b_{2} + b_{3}^{\dagger} b_{3} \right) \nonumber \\
&+ g \Big( \mathrm{e}^{\mathrm{i} \theta_{12}} b_{1}^{\dagger} b_{2} + \mathrm{e}^{\mathrm{i} \theta_{23}} b_{2}^{\dagger} b_{3} + \mathrm{e}^{\mathrm{i} \theta_{31}} b_{3}^{\dagger} b_{1} + \mathrm{h. c.} \Big),
\end{align}
where we have used cyclic boundary conditions, corresponding to the triangle geometry sketched in Fig.~\ref{sketch}~(a). The bosonic creation (annihilation) operators of harmonic oscillator $n$ are denoted by $b_n^\dagger$ ($b_n$), and each oscillator is associated with the resonance frequency $\omega_0$. The coherent coupling between oscillators $n$ and $m$ is of magnitude $g \ge 0$ and complex argument $\theta_{nm}$.

\begin{figure*}[tb]
 \includegraphics[width=1.0\linewidth]{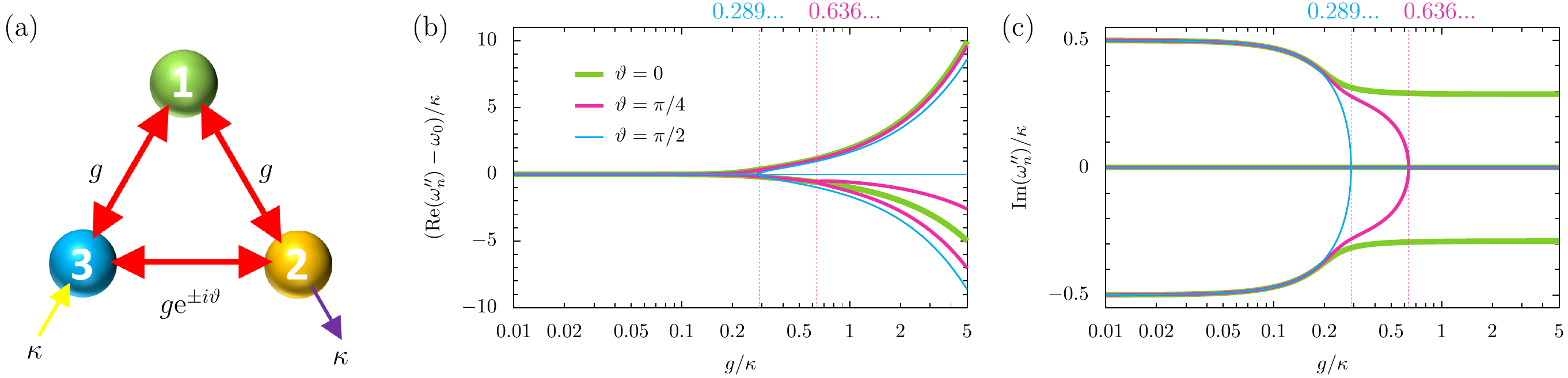}
 \caption{ Panel (a): a sketch of the $\mathcal{PT}$ symmetric trimer, where each oscillator is of resonance frequency $\omega_0$. The $2$nd oscillator is subject to loss $\kappa$ (purple arrow) and the $3$rd oscillator to gain $\kappa$ (yellow arrow). The magnitude of the three coherent coupling constants is $g$, and the hopping between oscillator $2$ and $3$ is augmented with the complex argument $\vartheta$. Panel (b): the real parts of the eigenfrequencies $\omega_n''$ of the trimer, measured from $\omega_0$, as a function of $g$, in units of $\kappa$ [cf. Eq.~\eqref{eq:Eig_one_PT}]. We consider three increasingly large phases $\vartheta = \{0, \pi/4, \pi/2 \}$, which are denoted by increasingly thin lines. Panel (c): the corresponding imaginary parts of the eigenfrequencies $\omega_n''$. Vertical dashed lines: exceptional points marking broken-unbroken $\mathcal{PT}$ symmetry phases. }
 \label{PTsketch}
\end{figure*}

Application of the Heisenberg equation of motion $[\beta_n, \hat{H}] = \omega_n \beta$, with Eq.~\eqref{eq:Ham} and $\beta_n = A_n b_1 + B_n b_2 + C_n b_3$, where $A_n, B_n, C_n$ are unknown constants, leads to the following matrix
\begin{equation}
\label{eq:Ham_one}
H =
\begin{pmatrix}
\omega_{0} && g \mathrm{e}^{-\mathrm{i} \theta_{12}} && g \mathrm{e}^{\mathrm{i} \theta_{31}} \\ 
g \mathrm{e}^{\mathrm{i} \theta_{12}} && \omega_{0} && g \mathrm{e}^{-\mathrm{i} \theta_{23}} \\ 
g \mathrm{e}^{-\mathrm{i} \theta_{31}} && g \mathrm{e}^{\mathrm{i} \theta_{23}} && \omega_{0} 
 \end{pmatrix},
\end{equation}
whose eigenvalues are the three eigenfrequencies $\omega_n$ of the system
\begin{subequations}
\label{eq:Eig_one}
\begin{alignat}{3}
\omega_{1} &= \omega_0 + 2 g \cos \left( \frac{\Theta + 2 \pi}{3} \right), \\
\omega_{2} &= \omega_0 + 2 g \cos \left( \frac{\Theta+ 4 \pi}{3} \right), \\
\omega_{3} &= \omega_0 + 2 g \cos \left( \frac{\Theta}{3} \right),
\end{alignat}
\end{subequations}
which lie within the range $\omega_0 -2g \le \omega_n \le \omega+2g$. In Eq.~\eqref{eq:Eig_one}, we have introduced the quantity
\begin{equation}
\label{eq:angle}
\Theta = \theta_{12} + \theta_{23} + \theta_{31},
\end{equation}
which describes the accumulated phase in the triangular closed loop of Fig.~\ref{sketch}~(a). We plot in Fig.~\ref{sketch}~(b) the eigenfrequencies $\omega_n$ of Eq.~\eqref{eq:Eig_one}, as a function of the cumulative phase $\Theta$ [cf. Eq.~\eqref{eq:angle}]. Most noticeably, $\Theta$ crucially determines the magnitude, ordering, and degeneracy of the eigenfrequencies. Notably, the trimer is the simplest system to exhibit such phase-dependent behavior. In the analogous dimer system, with Hamiltonian $H_{\mathrm{di}} = \omega_0 ( b_{1}^{\dagger} b_{1} + b_{2}^{\dagger} b_{2}) + g ( \mathrm{e}^{\mathrm{i} \theta_{12}} b_{1}^{\dagger} b_{2} + \mathrm{h.c.} )$, the resultant eigenfrequencies $\omega_{\pm}$ are independent of the complex argument $\theta_{12}$, and simply read $\omega_{\pm} = \omega_0 \pm g$. Hence the trimer is the perfect testbed to investigate phase-dependent phenomena.


\subsection*{Master equation}
\label{Master}

\begin{figure*}[tb]
 \includegraphics[width=0.75\linewidth]{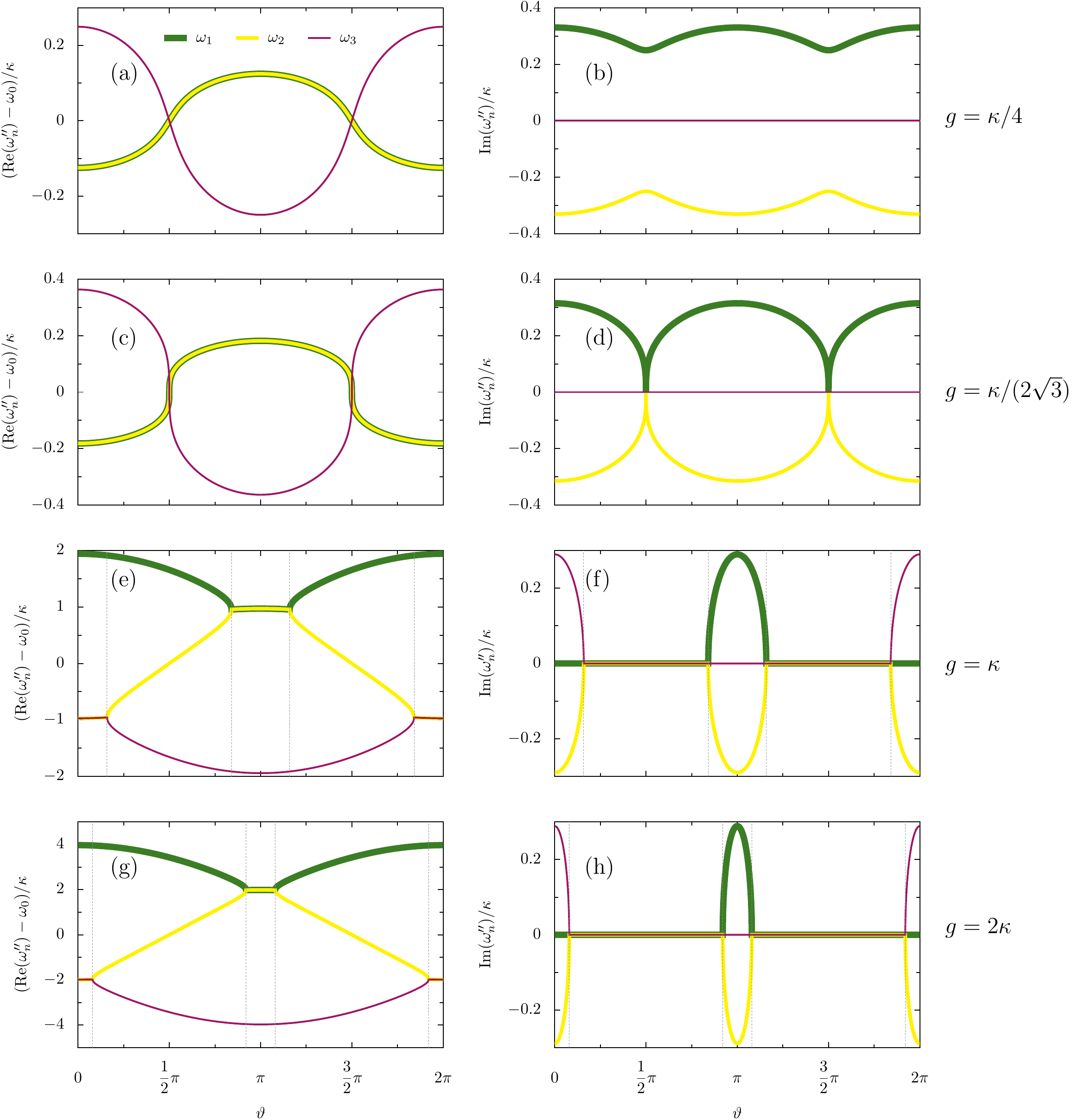}
 \caption{ Left-hand side panels: the real parts of the eigenfrequencies $\omega_n''$ [cf. Eq.~\eqref{eq:Eig_one_PT}], measured from $\omega_0$, as a function of the phase $\vartheta$ in the $\mathcal{PT}$ symmetric setup of the trimer [cf. Fig.~\ref{PTsketch}~(a)]. We consider increasing coherent coupling strength $g$, in units of the gain and loss parameter $\kappa$, descending the column of panels. Right-hand side panels: the corresponding imaginary parts of the eigenfrequencies $\omega_n''$. Vertical gray dashed lines: exceptional points marking broken-unbroken $\mathcal{PT}$ symmetry phases. }
 \label{PTeig}
\end{figure*}

\begin{figure*}[tb]
 \includegraphics[width=1.0\linewidth]{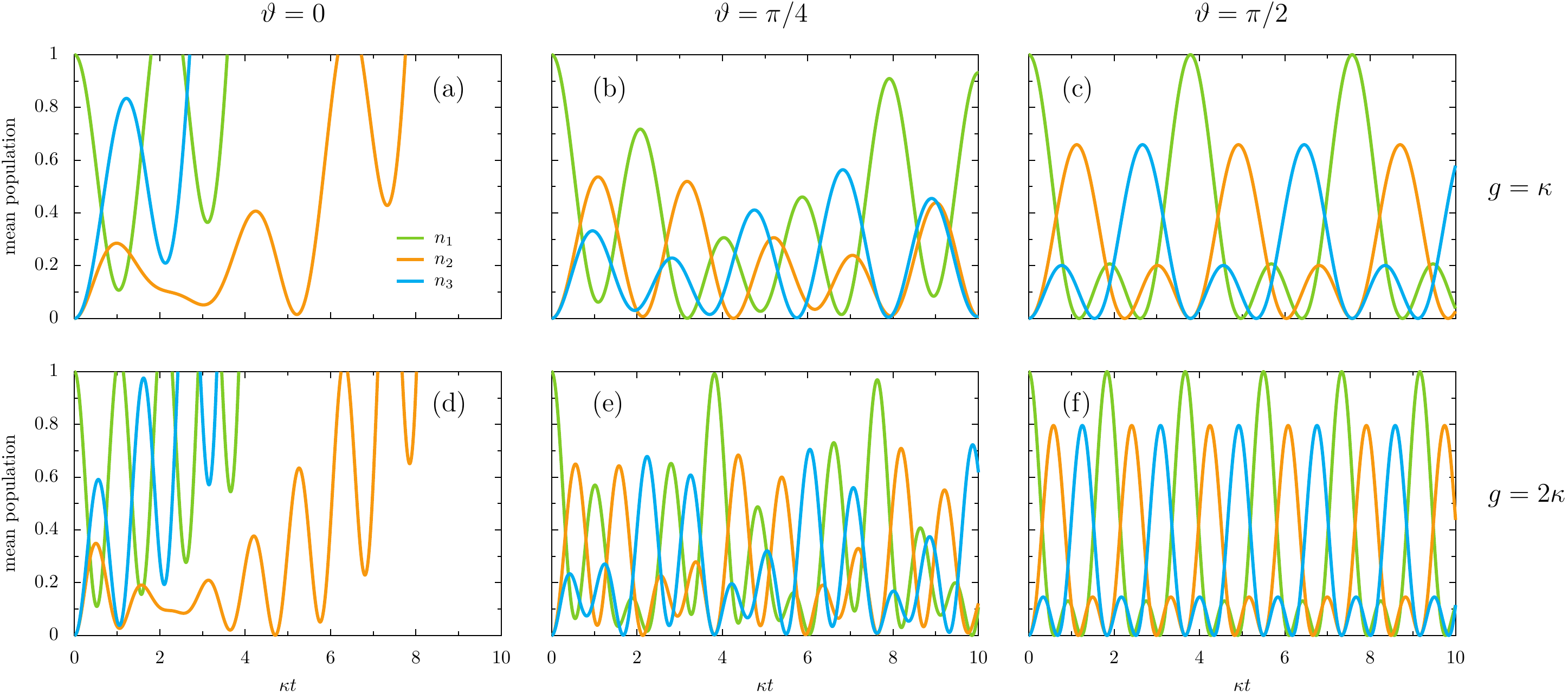}
 \caption{ Mean population $n_m = \langle b_m^\dagger b_m \rangle$ of the $m$-th oscillator as a function of time $t$, in units of the inverse of the gain and loss parameter $\kappa^{-1}$, in the $\mathcal{PT}$ symmetric setup of the trimer [cf. Fig.~\ref{PTsketch}~(a)]. We consider increasingly large phases $\vartheta$ across each row of panels, from $0$ to $\pi/4$ to $\pi/2$. The initial condition at $t=0$ is $n_1 = 1$. Upper panels: the coherent coupling strength $g = \kappa$. Lower panels: $g = 2\kappa$. }
 \label{PTpop}
\end{figure*}

The quantum master equation of the trimer system reads
\begin{align}
\label{eq:master}
 \partial_t \rho =&~\mathrm{i} [ \rho, \hat{H} ] 
 + \sum_{ n = 1, 2, 3 } \frac{P_n}{2} \mathcal{L}_n^{\dagger} \rho \nonumber \\
&+ \sum_{ n = 1, 2, 3 } \frac{\gamma_{n}}{2} \mathcal{L}_n \rho
+  \sum_{\substack{n, m = 1, 2, 3\\
                  n \ne m}}
                  \frac{\gamma}{2} \mathrm{e}^{\mathrm{i} \phi_{n m}} \mathcal{L}_{nm} \rho,
\end{align}
where we have used the following super-operators in Lindblad form~\cite{Gardiner2014}
\begin{subequations}
\label{eq:master2}
\begin{alignat}{3}
 \mathcal{L}_n \rho &= 2 b_n \rho b_n^{\dagger} -  b_n^{\dagger} b_n \rho - \rho b_n^{\dagger} b_n, \\
  \mathcal{L}_n^{\dagger} \rho &= 2 b_n^{\dagger} \rho b_n -  b_n b_n^{\dagger} \rho - \rho b_n b_n^{\dagger}, \\
 \mathcal{L}_{nm} \rho &= 2 b_{m} \rho b_n^{\dagger} -  b_n^{\dagger} b_{m} \rho - \rho b_n^{\dagger} b_{m}.
 \end{alignat}
\end{subequations}
In Eq.~\eqref{eq:master}, the first term on the right-hand-side is responsible for the unitary evolution, where the Hamiltonian operator $\hat{H}$ is given by Eq.~\eqref{eq:Ham}. The second term describes incoherent gain processes, where $P_n \ge 0$ is the pumping rate into the $n$-th oscillator. The third term accounts for losses into the heat bath, where $\gamma_n \ge 0$ is the damping decay rate of each individual oscillator. The fourth and final term describes the collective process of dissipative coupling, where the magnitude of the dissipative coupling is $\gamma \ge 0$. Each dissipative coupling constant is associated with a complex argument $\{ \phi_{12}, \phi_{23}, \phi_{31} \}$, in exactly the same manner as for the coherent coupling constants $g \, \mathrm{exp} ( \mathrm{i} \theta_{nm})$ appearing in Eq.~\eqref{eq:Ham}. Due to the necessity for Hermicity, the identities $\phi_{21} =  -\phi_{12}$, $ \phi_{32} =  -\phi_{23}$ and $ \phi_{13} =  -\phi_{31}$ are implied in the final term of Eq.~\eqref{eq:master}. The phase factors $\phi_{nm}$ associated with the incoherent couplings are analogues of the Rabi phases $\theta_{nm}$ accompanying the coherent couplings. Indeed, a gauge transformation can move these phase factors into the coherent coupling part of the Hamiltonian, so that it is actually the phase difference $\theta_{nm}-\phi_{nm}$ which is important [as we shall see later on in Eq.~\eqref{eq:casc_cond}]. The interplay between coherent and incoherent couplings has recently been shown to be influential for cavity-cavity interactions in photonic devices~\cite{Metelmann2015}, and for emitter-emitter interactions between circularly polarized quantum emitters~\cite{Downing2019}.

Taken together, Eq.~\eqref{eq:Ham} and Eq.~\eqref{eq:master} complete the general formalism of our theory. As we shall see, our model includes various special cases of interest, including for $\mathcal{PT}$ symmetric quantum mechanics~\cite{Bender1998, Bender2018} when the gain and loss is balanced, and for cascaded quantum systems~\cite{Gardiner1993, Carmichael1993} under certain conditions of the dissipative coupling.


\subsection*{Mean populations}
\label{Meanpop}

The master equation of Eq.~\eqref{eq:master}, and the property $\expect{\mathcal{O}} = \Tr{ \left( \mathcal{O} \rho \right) }$ for any operator $\mathcal{O}$, leads to the following equation of motion for the mean values of the populations 
\begin{equation}
\label{eqapp:of_motion}
\frac{\mathrm{d}}{\mathrm{d} t} \mathbf{u}  =  \mathbf{P} - \mathbf{M} \mathbf{u},
\end{equation}
for the 9-vector of correlators $\mathbf{u}$, where
\begin{equation}
\label{eqapp:umatrix}
\mathbf{u} =
\begin{pmatrix}
  \mathbf{u}_1  \\
  \mathbf{u}_2  \\
  \mathbf{u}_2^{\dagger}
 \end{pmatrix}, \quad
 \mathbf{u}_1 =
\left(\begin{array}{c}
  \langle b_1^{\dagger} b_1 \rangle  \\
  \langle b_2^{\dagger} b_2 \rangle  \\
  \langle b_3^{\dagger} b_3 \rangle  
    \end{array}
\right), \quad
 \mathbf{u}_2 =
\left(\begin{array}{c}
 \langle b_1^{\dagger}  b_2 \rangle  \\
  \langle b_2^{\dagger}  b_3 \rangle   \\
  \langle b_3^{\dagger}  b_1 \rangle  
    \end{array}
\right).
 \end{equation}
The drive term $\mathbf{P}$, and the matrix of second moments $\mathbf{M}$, read
\begin{equation}
\label{eqapp:Pdrive}
 \mathbf{P} = 
\begin{pmatrix}
  P_1  \\
  P_2 \\
  P_3 \\
  \boldsymbol{0}_{6}
 \end{pmatrix}, \quad
\mathbf{M} = \begin{pmatrix}
  \mathbf{M}_{11} && \mathbf{M}_{12} && \mathbf{M}_{12}^{\ast}  \\
  \mathbf{M}_{21} && \mathbf{M}_{22} && \mathbf{M}_{23}   \\
  \mathbf{M}_{21}^{\ast} && \mathbf{M}_{23}^{\ast} && \mathbf{M}_{22} 
 \end{pmatrix},
 \end{equation}
where $\text{\bf{0}}_{n}$ is the zero matrix (of a single column and $n$-rows). In Eq.~\eqref{eqapp:Pdrive}, the off-diagonal sub-matrices of $\mathbf{M}$ are defined by
\begin{subequations}
\label{eqapp:umatrix22}
\begin{alignat}{3}
\mathbf{M}_{12} &=
\begin{pmatrix}
  \tilde{g}_{12} && 0 && \tilde{f}_{31} \\
  \tilde{f}_{12} && \tilde{g}_{23} && 0  \\
 0 && \tilde{f}_{23} && \tilde{f}_{31}
 \end{pmatrix}, \\  
 \mathbf{M}_{21} &=
\begin{pmatrix}
  \tilde{f}_{12}^{\ast} && \tilde{g}_{12}^{\ast} && 0 \\
  0 && \tilde{f}_{23}^{\ast} && \tilde{g}_{23}^{\ast}  \\
 \tilde{g}_{31}^{\ast} && 0 && \tilde{f}_{31}^{\ast}
 \end{pmatrix}, \\
 \mathbf{M}_{23} &=
\begin{pmatrix}
  0 && \tilde{f}_{31} && \tilde{g}_{23} \\
  \tilde{g}_{31} && 0 && \tilde{f}_{12}  \\
 \tilde{f}_{23} && \tilde{g}_{12} && 0
 \end{pmatrix},
 \end{alignat}
\end{subequations}
 where we have introduced the generalized coupling constants
 \begin{align}
\label{eqapp:fg22}
 \tilde{g}_{nm} &= \mathrm{i} g \mathrm{e}^{\mathrm{i} \theta_{nm}} + \frac{\gamma}{2} \mathrm{e}^{\mathrm{i} \phi_{nm}}, \\
 \tilde{f}_{nm} &= -\mathrm{i} g \mathrm{e}^{\mathrm{i} \theta_{nm}} + \frac{\gamma}{2} \mathrm{e}^{\mathrm{i} \phi_{nm}},
 \end{align}
 which take into account both coherent $g \, \mathrm{exp} ( \mathrm{i} \theta_{nm})$ and dissipative $\gamma \, \mathrm{exp} ( \mathrm{i} \phi_{nm})$ couplings in the trimer. Finally, the on-diagonal sub-matrices comprising $\mathbf{M}$ are given by
 \begin{subequations}
\label{eqapp:umatrasasix22}
\begin{alignat}{3}
\mathbf{M}_{11} &= \mathrm{diag} \left( \Gamma_1, \Gamma_2, \Gamma_3 \right), \\
 \mathbf{M}_{22} &= \mathrm{diag} \left(  \tfrac{\Gamma_1+\Gamma_2}{2}, \tfrac{\Gamma_2+\Gamma_3}{2}, \tfrac{\Gamma_3+\Gamma_1}{2} \right). 
 \end{alignat}
 \end{subequations}
Here the renormalized damping decay rate $\Gamma_n$ of the $n$-th oscillator, due to the incoherent pumping $P_n$, is
 \begin{equation}
\label{eqapp:ffsdfsg22}
 \Gamma_n = \gamma_n - P_n.
 \end{equation}
The formal solution of Eq.~\eqref{eqapp:of_motion} leads to the mean populations $n_m = \langle b_m^\dagger b_m \rangle$ for the $m$-th oscillator in the trimer, which is in general composed of both transient and steady state parts.

\subsection*{Current}
\label{app:current}

\begin{figure*}[tb]
 \includegraphics[width=0.75\linewidth]{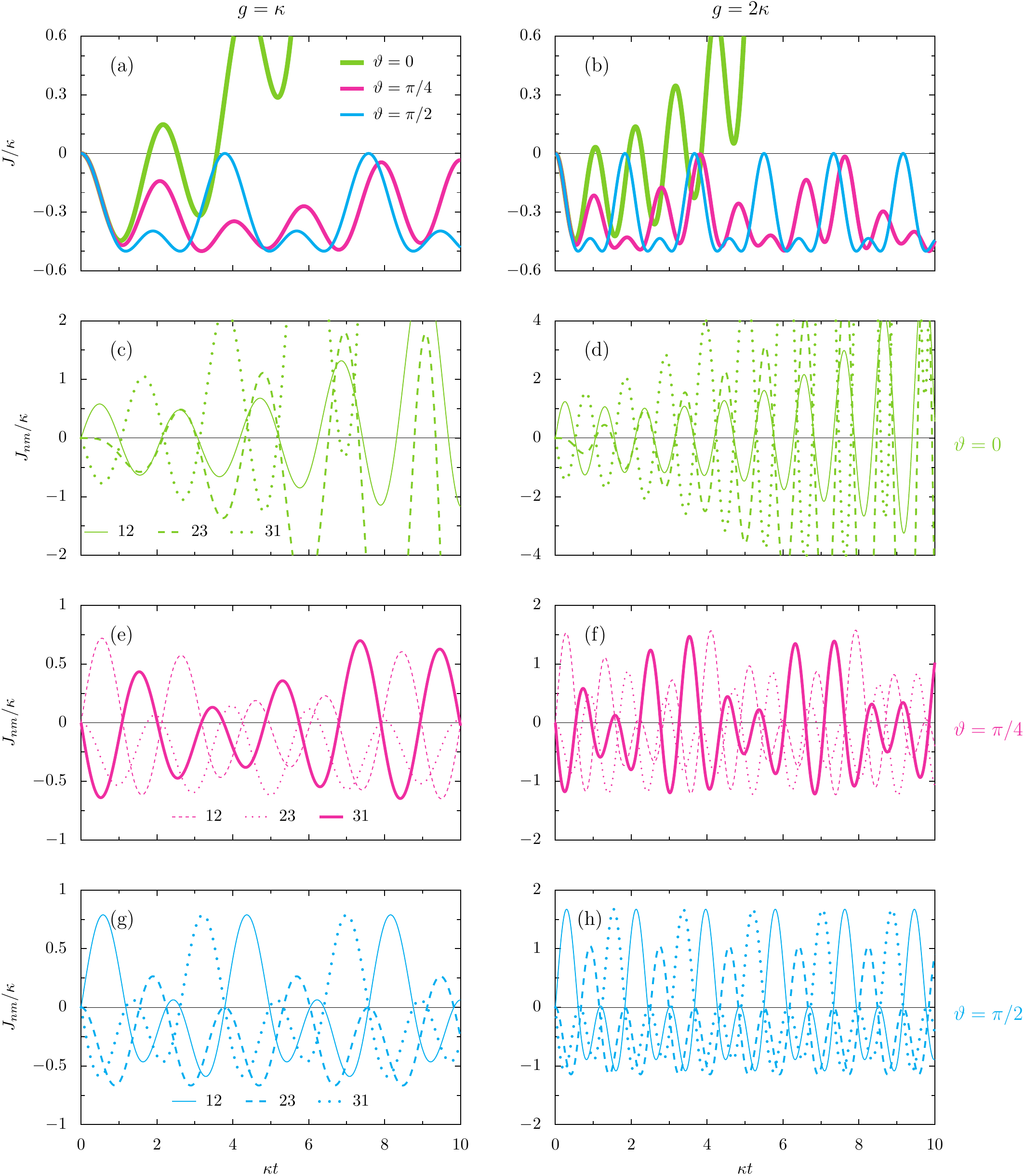}
 \caption{ Top row: Global current $J$ [cf. Eq.~\eqref{eqapp:current4}], in units of the gain and loss parameter $\kappa$, as a function of time $t$, in the $\mathcal{PT}$ symmetric setup of the trimer [cf. Fig.~\ref{PTsketch}~(a)]. We consider increasingly large phases $\vartheta$ with increasingly thin colored lines, from $0$ to $\pi/4$ to $\pi/2$. Lower rows: scaled local currents $J_{nm}/\kappa$, as a function of scaled time $\kappa t$, for the three phases $\vartheta$ corresponding to the top panels [cf. Eq.~\eqref{eqapp:current4}]. The solid, dashed and dotted lines represent $J_{12}$, $J_{23}$, and $J_{31}$ respectively. Left-hand-side column: the coherent coupling strength $g = \kappa$. Right-hand-side column: $g = 2\kappa$. In  the figure, the initial condition at $t=0$ is $n_1 = 1$.  }
 \label{PTcurrent}
\end{figure*}

Here we derive an expression for the current flowing around the three site loop, as depicted in Fig.~\ref{sketch}~(a). The continuity equation at each site $n$ in the trimer reads
\begin{equation}
\label{eqapp:current1}
 \partial_t \left( b_n^\dagger b_n \right) = \mathrm{i} [ b_n^\dagger b_n , \hat{H} ] = I_{n n+1} - I_{n-1 n},
 \end{equation}
 where the Hamiltonian operator $\hat{H}$ is given by Eq.~\eqref{eq:Ham}. In Eq.~\eqref{eqapp:current1}, we have introduced the local current operator $I_{n n+1}$ which accounts for the transfer of excitations between two adjacent sites $n$ and $n+1$ (we assume modular arithmetic for the site numbers), 
\begin{equation}
\label{eqapp:current2}
      I_{n n+1} = \mathrm{i} g \left( \mathrm{e}^{\mathrm{i} \theta_{n n+1}} b_{n}^{\dagger}b_{n+1}  - \mathrm{e}^{-\mathrm{i} \theta_{n n+1}} b_{n+1}^{\dagger}b_{n} \right).
\end{equation} 
Similarly, the global current operator can be defined as
 \begin{equation}
\label{eqapp:current3}
 I = I_{12} + I_{23} + I_{31}.
 \end{equation}
Notably, the current definition of Eq.~\eqref{eqapp:current1} does not explicitly depend on the loss and gain in the system, instead these features enter the consideration indirectly through the operators $b_n^\dagger b_m$. We denote the mean versions of the quantities given by Eq.~\eqref{eqapp:current2} and Eq.~\eqref{eqapp:current3} as
 \begin{equation}
\label{eqapp:current4}
  J = \langle I \rangle, \quad J_{n n+1} = \langle I_{n n+1} \rangle,
 \end{equation}
which may be explicitly calculated using the mean values of the operators $\langle b_{n}^{\dagger} b_{m} \rangle$, as found previously.


\section*{$\mathcal{PT}$ symmetric trimer}
\label{PTsymm}

\begin{figure}[tb]
 \includegraphics[width=0.55\linewidth]{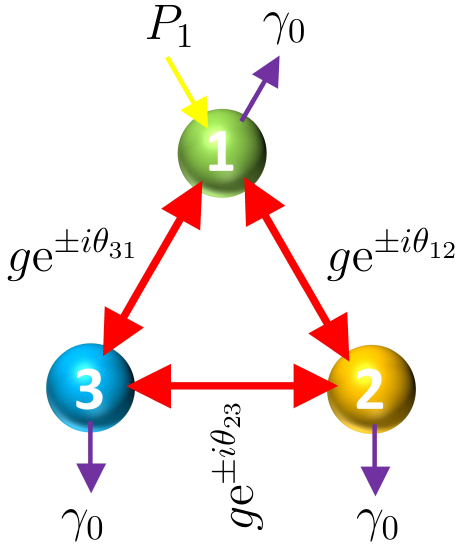}
 \caption{ A sketch of the trimer in a chiral configuration, where each harmonic oscillator is of resonance frequency $\omega_0$. All oscillators are subject to loss $\gamma_0$ (purple arrows) and the $1$st oscillator has gain $P_1$ (yellow arrow). The magnitude of the three coherent coupling constants is $g$, which are augmented with complex arguments $\theta_{nm}$, as labelled near to the red arrows. }
 \label{chiralsketch}
\end{figure}

\begin{figure*}[tb]
 \includegraphics[width=1.0\linewidth]{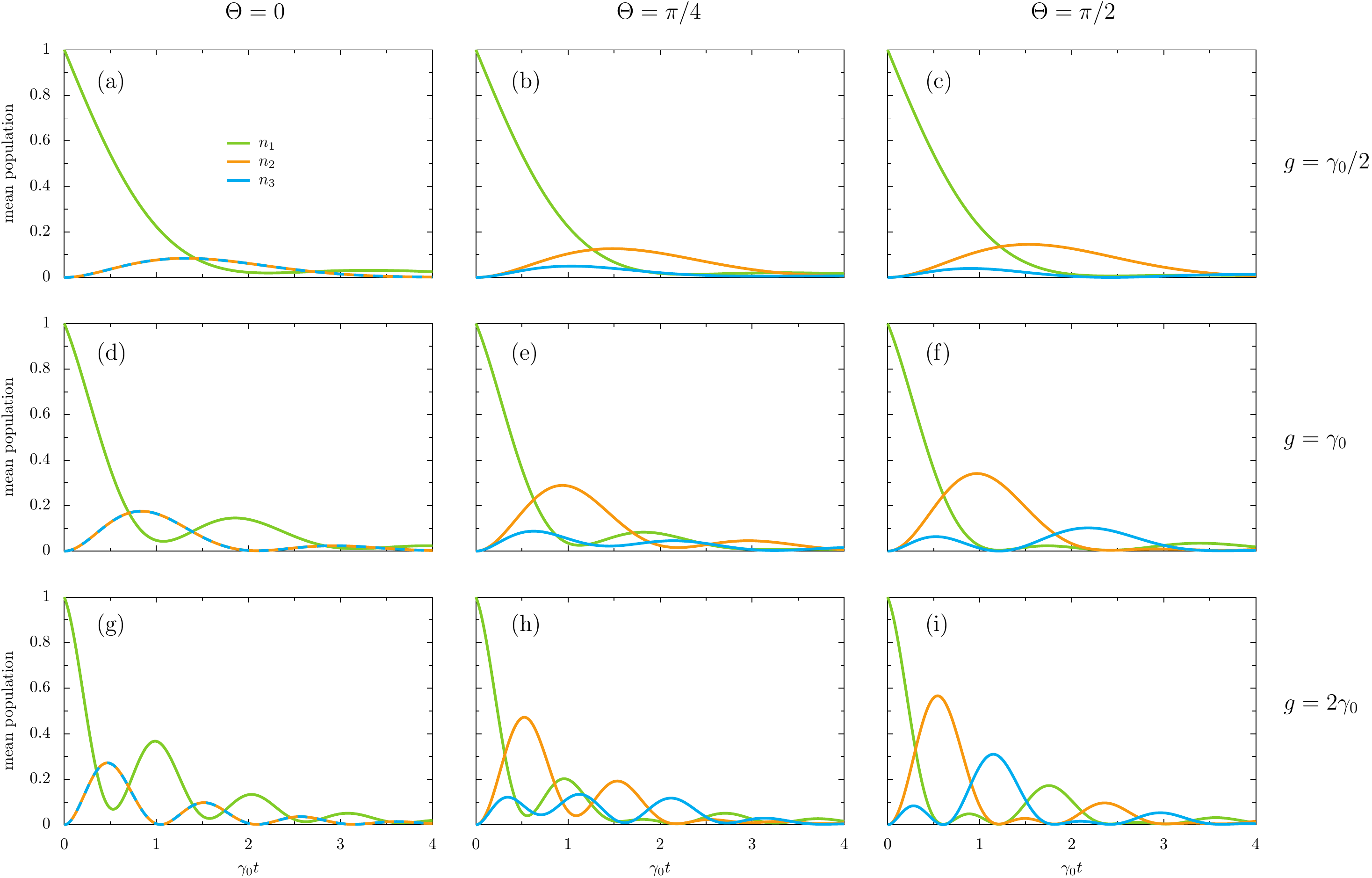}
 \caption{ Mean population $n_m = \langle b_m^\dagger b_m \rangle$ of the $m$-th harmonic oscillator as a function of time $t$, in units of the inverse of the loss parameter $\gamma_0^{-1}$, in the chiral configuration of the trimer [cf. Fig.~\ref{chiralsketch}]. We consider increasingly large cumulative phases $\Theta$ across each row of panels, from $0$ to $\pi/4$ to $\pi/2$. The initial condition at $t=0$ is $n_1 = 1$. Upper panels: the coherent coupling strength $g = \gamma_0/2$. Middle panels: $g = \gamma_0$. Lower panels: $g = 2\gamma_0$. In the figure, the gain into oscillator $1$ is $P_1 = \gamma_0/100$ and $P_2 = P_3 = 0$. }
 \label{chiralpop}
\end{figure*}

As a first application of the general theory, we consider a $\mathcal{PT}$ symmetric setup of the trimer. To do so, we first simplify our notation from Eq.~\eqref{eq:Ham} and consider the relabeled Hamiltonian
\begin{align}
\label{eq:Ham_PT}
\hat{H'} =& ~\omega_0 \left( b_{1}^{\dagger} b_{1}  + b_{2}^{\dagger} b_{2} + b_{3}^{\dagger} b_{3} \right) \nonumber \\
&+ g \Big( b_{1}^{\dagger} b_{2} + \mathrm{e}^{\mathrm{i} \vartheta} b_{2}^{\dagger} b_{3} +  b_{3}^{\dagger} b_{1} + \mathrm{h. c.} \Big),
\end{align}
where we have allowed for a complex argument $\vartheta$, associated with the hopping between the oscillators $2$ and $3$. As sketched in Fig.~\ref{PTsketch}~(a), the $2$nd oscillator is subject to loss via the parameter $\kappa$ (purple arrow) and the $3$rd oscillator to gain at the same rate $\kappa$ (yellow arrow). Such a scenario can be described by the Hamiltonian~\cite{Supporting}
\begin{equation}
\label{eq:Ham_PT2}
\hat{H''} = \hat{H'} - \mathrm{i} \frac{\kappa}{2} b_{2}^{\dagger} b_{2} + \mathrm{i} \frac{\kappa}{2} b_{3}^{\dagger} b_{3},
\end{equation}
in terms of $\hat{H'}$ from Eq.~\eqref{eq:Ham_PT}.

The Hamiltonian $\hat{H''}$ is non-Hermitian ($\hat{H''} \ne \hat{H''}^\dagger$) but it can nevertheless have wholly real eigenvalues, as was first realized by Bender and co-workers~\cite{Bender1998}, since Hermicity is not a necessary condition for a physical quantum theory~\cite{Bender2018}. Importantly, the Hamiltonian of Eq.~\eqref{eq:Ham_PT2} is $\mathcal{PT}$ symmetric $[PT, \hat{H''}] = 0$, where $P$ and $T$ are parity inversion and time reversal symmetry operators~\cite{mynoto}. Therefore, in its unbroken phase, Eq.~\eqref{eq:Ham_PT2} will produce real eigenvalues, while its broken phase is associated with complex eigenvalues. Explicitly, the three complex eigenfrequencies $\omega_{i}''$ of $\hat{H''}$ read
\begin{subequations}
\label{eq:Eig_one_PT}
\begin{alignat}{3}
\omega_{1}'' &= \omega_0 + 2 \sqrt{g^2 - \frac{1}{3} \left( \frac{\kappa}{2} \right)^2  } \cos \left( \frac{\alpha + 2 \pi}{3} \right), \\
\omega_{2}'' &= \omega_0 + 2 \sqrt{g^2 - \frac{1}{3} \left( \frac{\kappa}{2} \right)^2  } \cos \left( \frac{\alpha+ 4 \pi}{3} \right), \\
\omega_{3}'' &= \omega_0 + 2 \sqrt{g^2 - \frac{1}{3} \left( \frac{\kappa}{2} \right)^2  } \cos \left( \frac{\alpha}{3} \right),
\end{alignat}
\end{subequations}
where we have introduced the angle
\begin{equation}
\label{eq:Ham_PT2_b}
\cos \left( \alpha \right) = \frac{\cos \left( \vartheta \right)}{ \left[ 1 - \frac{1}{3} \left( \frac{\kappa}{2 g} \right)^2 \right]^{3/2} }.
\end{equation}
Clearly, in the Hermitian limit of $H''$ (when $\kappa \to 0$) the eigenfrequencies of Eq.~\eqref{eq:Eig_one_PT} recover those of Eq.~\eqref{eq:Eig_one}, after the relabeling of the phase $\vartheta \to \Theta$. In the simple case where $\vartheta = \pi/2$, Eq.~\eqref{eq:Eig_one_PT} reduces to the three eigenfrequencies $\omega_0 \pm \sqrt{3 g^2 - (\kappa/2)^2}$ and $\omega_0$. Most notably, these expressions reveal completely real eigenvalues when the coherent coupling strength $g > \kappa / (2\sqrt{3})$, and otherwise a broken phase when $g < \kappa / (2\sqrt{3})$, with an exceptional point at $g = \kappa / (2\sqrt{3}) \simeq 0.289 \kappa$.

We plot the real and imaginary parts of the three eigenfrequencies $\omega_n''$ from Eq.~\eqref{eq:Eig_one_PT} in Fig.~\ref{PTsketch} panels (b) and (c) respectively, as a function of the coherent coupling strength $g$. We select three increasingly large phases of interest $\vartheta = \{0, \pi/4, \pi/2 \}$, which are denoted by increasingly thin lines. The broken-unbroken phase transitions at exceptional points are shown by the vertical dashed lines, including the aforementioned simple case when $\vartheta = \pi/2$ (thin cyan lines). A similar behavior is displayed when $\vartheta = \pi/4$ (medium pink lines), where the critical coupling strength is increased to $g \simeq 0.636 \kappa$. However, when $\vartheta = 0$ (thick green lines) the system is always in a broken $\mathcal{PT}$ phase, as is apparent from panel (c). Hence, both the interplay between the ratio of the system parameters $g/\kappa$ and the phase $\vartheta$ crucially determine the nature of the $\mathcal{PT}$ symmetry (either broken or unbroken) associated with the non-Hermitian Hamiltonian $H''$, as given by Eq.~\eqref{eq:Ham_PT2}.

We examine the $\mathcal{PT}$ phase transitions in more detail in Fig.~\ref{PTeig}. The left-hand side panels show the real parts of the eigenfrequencies $\omega_n''$ as a function of the phase $\vartheta$, while the right-hand side panels display the corresponding imaginary parts of the eigenfrequencies [cf. Eq.~\eqref{eq:Eig_one_PT}]. The first row of panels~(a, b) shows the situation with weak coupling $g = \kappa/4$, where two eigenfrequencies $\omega_n''$ are complex for all values of $\vartheta$ (green and yellow lines). This broken $\mathcal{PT}$ phase continues with increasing coherent coupling strength, until the critical point $g = \kappa / (2 \sqrt{3})$ [in the second row of panels~(c, d)]. Here $1 / (2 \sqrt{3}) \simeq 0.289$ is a universal constant describing the threshold above which unbroken $\mathcal{PT}$ phases may first arise, for at least some values of $\vartheta$. The third row of panels~(e, f), where $g = \kappa$, demonstrates that above this critical condition phases of broken and unbroken $\mathcal{PT}$ symmetry are both possible, as marked by the vertical dashed lines at the exceptional points. With further increased coupling strength, like when $g = 2\kappa$ in the bottom row of panels~(g, h), the regions of broken $\mathcal{PT}$ symmetry are increasingly diminished, and become increasingly concentrated around the reciprocal phases of $\vartheta = \{ 0, \pi, 2\pi \}$. This analysis highlights the rich physics exhibited by the trimer $\mathcal{PT}$ Hamiltonian of Eq.~\eqref{eq:Ham_PT2}, and the crucial importance of the phase $\vartheta$. Notably, for a simple dimer $\mathcal{PT}$ model the critical coupling strength is fixed at $g = \kappa/2$, independent of any accumulated phase~\cite{Ganainy2018, Ozdemir2019}.

Armed with the knowledge of the regions of real and complex eigenfrequencies $\omega_n''$, we investigate the population dynamics in the $\mathcal{PT}$ symmetric setup of the trimer. We use the transient solutions of Eq.~\eqref{eqapp:of_motion}, arising from the theoretical framework developed previously. The results are presented in Fig.~\ref{PTpop}, where we show the mean population $n_m = \langle b_m^\dagger b_m \rangle$ of the $m$-th harmonic oscillator as a function of time $t$, in units of the inverse gain and loss parameter $\kappa^{-1}$. In the upper (lower) panels, the coherent coupling strength $g = \kappa$ ($g = 2\kappa$). We consider increasingly large phases $\vartheta$ across each row of panels, from $0$ to $\pi/4$ to $\pi/2$. In the first column of Fig.~\ref{PTpop}, in panels~(a)~and~(d) where $\vartheta = 0$, we are in a broken $\mathcal{PT}$ phase and as such the dynamics is unstable and divergent [cf. Fig.~\ref{PTeig}~(e, f)~and~(g, h) respectively]. Upon increasing the phase to $\vartheta = \pi/4$, as shown in the middle columns of panels~(b, e), we enter an unbroken $\mathcal{PT}$ phase, associated with stable population dynamics. Here the directional circulation in the trimer is most apparent, with the larger coherent coupling in the panel~(e) causing even higher mean populations. The final column of panels~(c, f), where $\vartheta = \pi/2$, displays highly directional circulation in a simple structural form: from oscillator $1$ (green lines) to oscillator $2$ (orange lines) to $3$ (cyan lines) in a clockwise manner [cf. Fig.~\ref{PTsketch}~(a)]. 

We plot the global current $J$ in the $\mathcal{PT}$ symmetric trimer in the top row of Fig.~\ref{PTcurrent}, using Eq.~\eqref{eqapp:current4}, which complements the physics presented in Fig.~\ref{PTpop}. We consider increasingly large cumulative phases $\Theta$ in Fig.~\ref{PTcurrent} with increasingly thin colored lines, from $0$ to $\pi/4$ to $\pi/2$. The lower rows display the constituent local currents $J_{12}$, $J_{23}$, and $J_{31}$, which are represented by solid, dashed and dotted lines respectively. In the left-hand-side column the coherent coupling strength $g = \kappa$, and in the right-hand-side column $g = 2\kappa$. The top row of Fig.~\ref{PTcurrent} highlights the broken $\mathcal{PT}$ phase when $\vartheta = 0$ (thick green lines), leading to divergent global currents $J$ in both panels (a) and (b). The cases in the unbroken $\mathcal{PT}$ phase with $\vartheta = \pi/4$ and $\pi/2$ (pink and cyan lines respectively) show regular behavior. The consistent global current amplitude minima and maxima in panels (a) and (b) arise due to the balanced loss and gain in the system. These global currents $J$ are not sign changing, as follows from the behavior of the local currents $J_{nm}$ through the sites $n$ and $m$, which are displayed in the lower panels for the three phases of interest.


\section*{Chiral currents}
\label{ChiralCurr}

\begin{figure*}[tb]
 \includegraphics[width=1.0\linewidth]{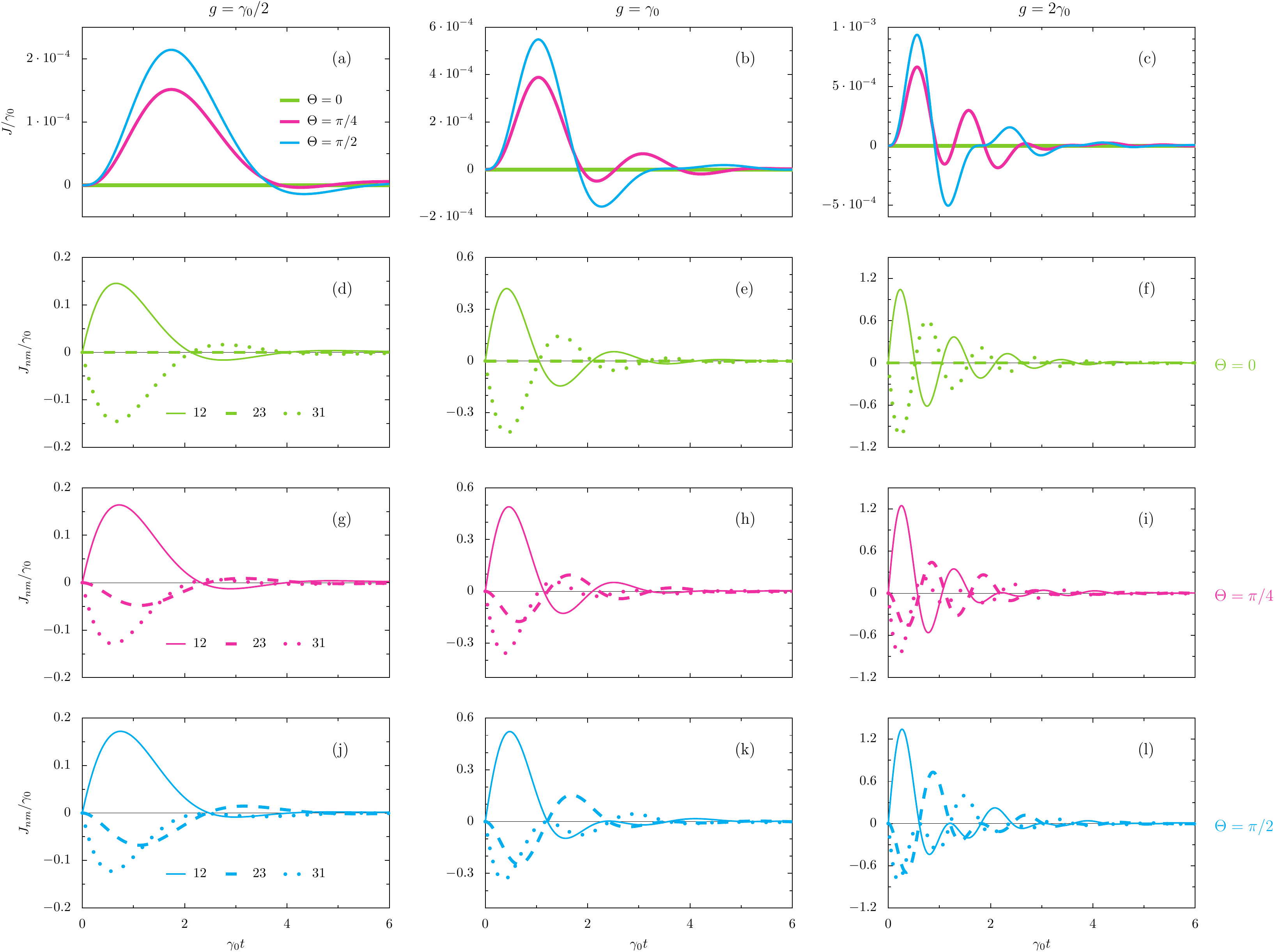}
 \caption{ Top row: Global current $J$ as a function of time $t$ [cf. Eq.~\eqref{eqapp:current4}], in units of the inverse of the loss parameter $\gamma_0^{-1}$, in the chiral configuration of the trimer [cf. Fig.~\ref{chiralsketch}]. We consider increasingly large cumulative phases $\Theta$ with increasingly thin colored lines, from $0$ to $\pi/4$ to $\pi/2$. Lower rows: scaled local currents $J_{nm}/\gamma_0$, as a function of scaled time $\gamma_0 t$, for the three phases $\Theta$ corresponding to the top panels [cf. Eq.~\eqref{eqapp:current4}]. The solid, dashed and dotted lines represent $J_{12}$, $J_{23}$, and $J_{31}$ respectively. Left-hand-side panels: the coherent coupling strength $g = \gamma_0/2$. Central panels: $g = \gamma_0$. Right-hand-side panels panels: $g = 2\gamma_0$. In the figure, the initial condition at $t=0$ is $n_1 = 1$, the gain into oscillator $1$ is $P_1 = \gamma_0/100$, and $P_2 = P_3 = 0$.   }
 \label{chiralcurrent}
\end{figure*}

As a second application of the general theory, we consider a chiral configuration of the trimer, beyond the rather restrictive $\mathcal{PT}$ setup of the previous section. As sketched in Fig.~\ref{chiralsketch}, we feed the system via the gain $P_1$ into the first oscillator (yellow arrow), and allow for equal losses $\gamma_0$ across all oscillators (purple arrows). Here we neglect dissipative coupling, and focus on the effects of purely coherent coupling, which is of strength $g$ and leads to an accumulated phase $\Theta$ in the triangular closed loop [cf. Eq.~\eqref{eq:angle}]. We calculate the mean populations and current in this configuration, using both the transient and steady state solutions of Eq.~\eqref{eqapp:of_motion}.

In Fig.~\ref{chiralpop}, we plot the mean population $n_m = \langle b_m^\dagger b_m \rangle$ of the $m$-th oscillator as a function of time $t$, where the gain into oscillator $1$ is $P_1 = \gamma_0/100$. We show increasingly large cumulative phases $\Theta$ across each row of panels, from $0$ to $\pi/4$ to $\pi/2$. Descending the columns allows one to see the effects of increased coherent coupling strengths $g$. We show results for $g = \gamma_0/2$ (top panels), $g = \gamma_0$ (middle) and $g = 2\gamma_0$ (bottom). In the first column of panels~(a, d, g) in Fig.~\ref{chiralpop}, where $\Theta = 0$, there is completely reciprocal dynamics, and the populations of the second (orange lines) and third (cyan lines) oscillators are identical. With nonzero accumulated phase $\Theta = \pi/4$ in the second column of panels~(b, e, h), a population imbalance is induced, as follows from the asymmetry in the population dynamics of the second and third oscillators (orange and cyan lines), and a few oscillation cycles are noticeable in panel (h). The final column of panels~(c, f, i), where $\Theta = \pi/2$, displays the strongest signature of directional circulation $1 \to 2 \to 3$ (or green $\to$ orange $\to$ cyan), which is most apparent when the coherent coupling is strongest in panel (i).

We study the global current $J$ in this setup in the top row of Fig.~\ref{chiralcurrent}, using Eq.~\eqref{eqapp:current4}. We consider increasingly large cumulative phases $\Theta$ with increasingly thin colored lines, from $0$ to $\pi/4$ to $\pi/2$. In the left-hand-side column, the coherent coupling strength $g = \gamma_0/2$, in the central column $g = \gamma_0$, and in the right-hand-side column $g = 2\gamma_0$. In the figure, the gain into oscillator $1$ is $P_1 = \gamma_0/100$. Figure~\ref{chiralcurrent} complements the population dynamics displayed in Fig.~\ref{chiralpop} by showing the absence of any global current $J$ when $\Theta = 0$ [thick green lines in panels (a-c)]. With a nontrivial phase $\Theta = \pi/4$ or $\pi/2$ (pink and cyan lines), a nonzero global current clearly emerges in all panels (a-c), before dying out at longer times $\gamma_0 t \gg 1$ (since the gain does not balance out against the losses, unlike the $\mathcal{PT}$ symmetric case).

In stark contrast to the $\mathcal{PT}$ symmetric setup of Fig.~\ref{PTcurrent}, here in Fig.~\ref{chiralcurrent} the global current $J$ is noticeably sign-changing for the nontrivial phases. This is because the component local currents $\{ J_{12}, J_{23}, J_{31} \}$ do not correspond to as strongly directional circulation. This behavior is displayed in the lower panels of Fig.~\ref{chiralcurrent}, where the solid, dashed and dotted cyan lines represent $J_{12}$, $J_{23}$, and $J_{31}$ respectively (in units of $\gamma_0$, and as a function of $\gamma_0 t$). In particular, the exact cancellations of $J_{12}$ and $J_{31}$ in the second row (d, e, f) lead to zero global current, a symmetry which is broken in the two lowest rows.

\begin{figure}[tb]
 \includegraphics[width=0.85\linewidth]{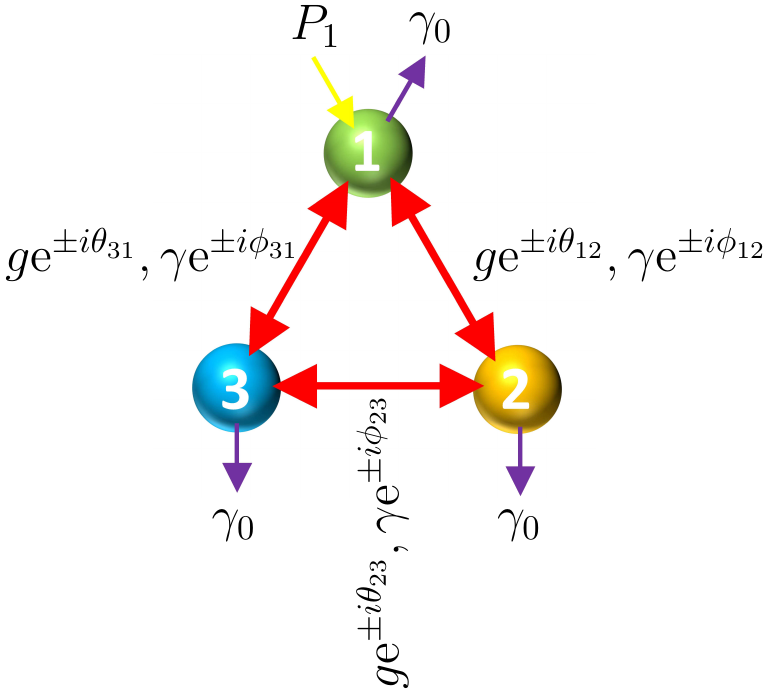}
 \caption{ A sketch of the trimer in a cascaded configuration, where each oscillator is of resonance frequency $\omega_0$. All oscillators are subject to loss $\gamma_0$ (purple arrows) and the $1$st oscillator has gain $P_1$ (yellow arrow). The magnitude of the three coherent (dissipative) coupling constants is $g$ ($\gamma$), which are augmented with complex arguments $\theta_{nm}$ ($\phi_{nm}$), as labelled near to the red arrows. }
 \label{cascadedsketch}
\end{figure}


\section*{Cascaded circulation}
\label{UniCurr}

\begin{figure*}[tb]
 \includegraphics[width=1.0\linewidth]{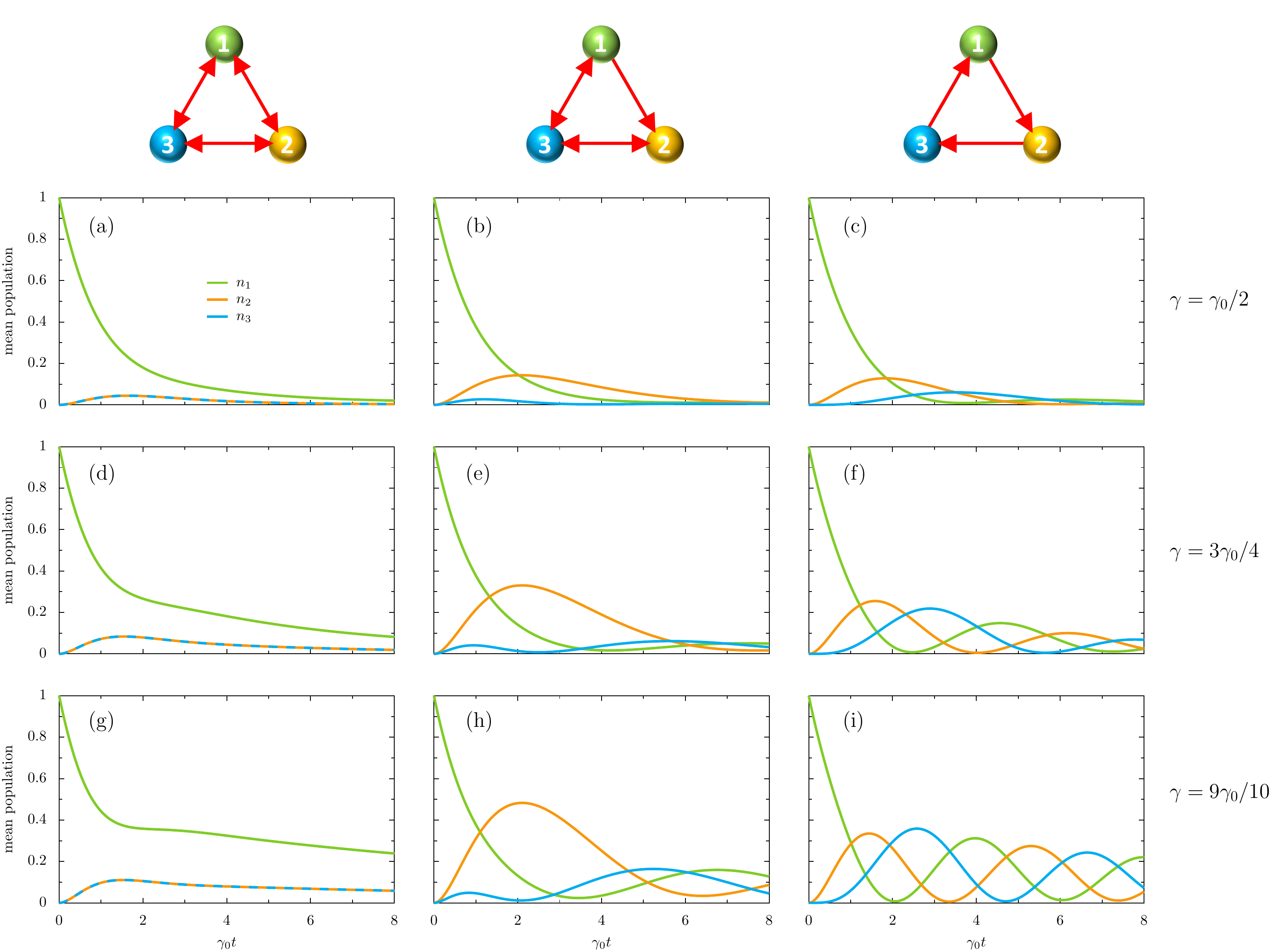}
 \caption{ Mean population $n_m = \langle b_m^\dagger b_m \rangle$ of the $m$-th oscillator as a function of time $t$, in units of the inverse loss parameter $\gamma_0^{-1}$, in the cascaded configuration of the trimer [cf. Fig.~\ref{cascadedsketch}]. The coherent coupling strength $g = \gamma /2$, in units of the dissipative coupling strength $\gamma$, satisfying the magnitude cascaded condition of Eq.~\eqref{eq:casc_cond_1}. The initial condition at $t=0$ is $n_1 = 1$. First column: the completely reciprocal case, where no complex arguments are attached to the coupling constants [see the sketch above panel~(a)]. Second column: a chiral case, where $\theta_{12} - \phi_{12} = \pi/2$, leads to cascaded coupling between oscillators $1$ and $2$ [see the sketch above panel~(b)]. Third column: the completely cascaded case [cf. Eq.~\eqref{eq:casc_cond_2}], where $\theta_{12} - \phi_{12} = \theta_{23} - \phi_{23} = \theta_{31} - \phi_{31} = \pi/2$, leading to clockwise circulation [see the sketch above panel~(c)]. We show increasingly strong dissipative coupling upon descending the rows, with $\gamma = \{ 1/2, 3/4, 9/10 \} \gamma_0$. In the figure, the gain into oscillator $1$ is $P_1 = \gamma_0/100$ and $P_2 = P_3  = 0$.  }
 \label{cascadedpop}
\end{figure*}

\begin{figure*}[tb]
 \includegraphics[width=1.0\linewidth]{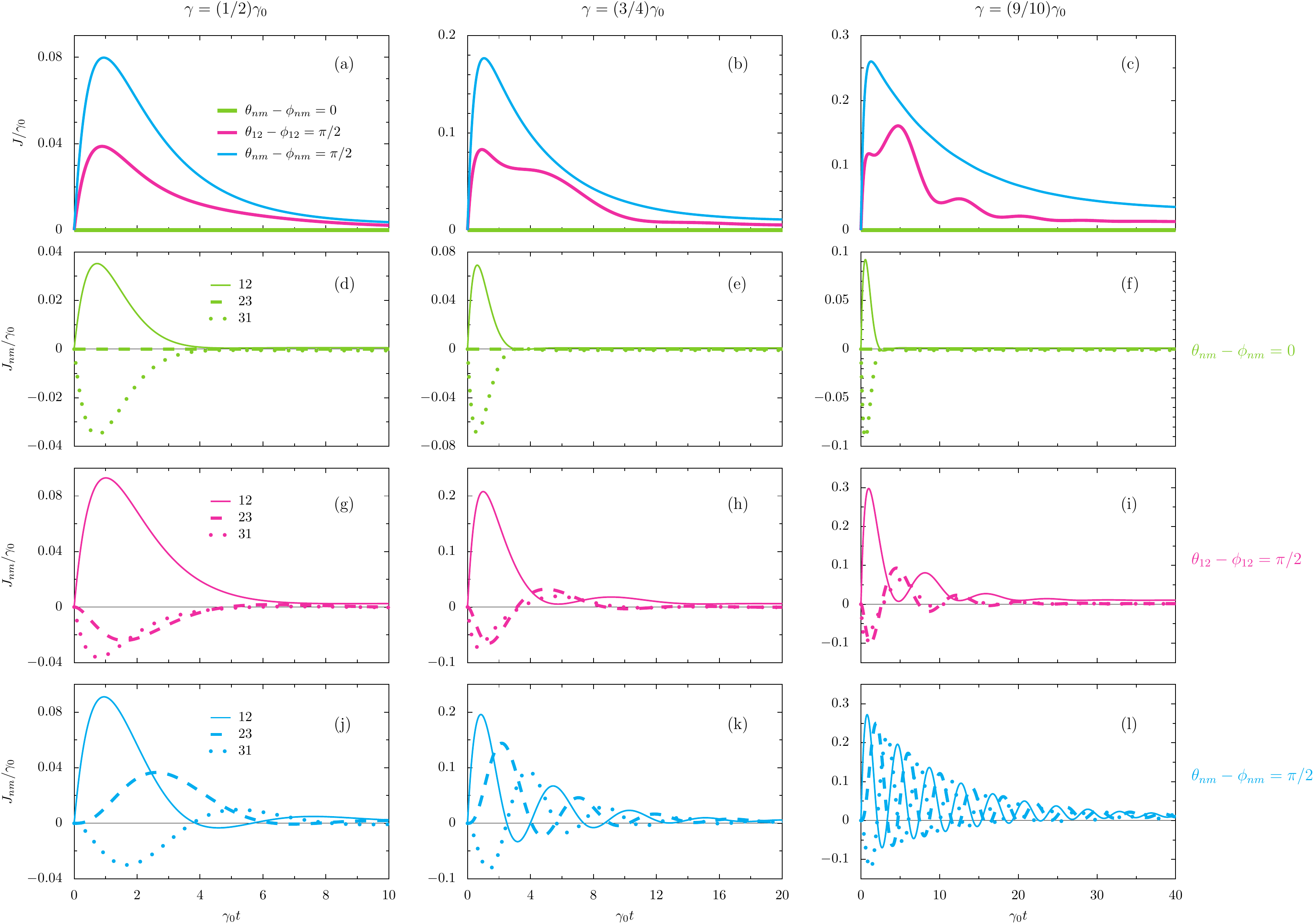}
 \caption{ Top row: Global current $J$ as a function of time $t$ [cf. Eq.~\eqref{eqapp:current4}], in units of the inverse loss parameter $\gamma_0^{-1}$, in the cascaded configuration of the trimer [cf. Fig.~\ref{cascadedsketch}]. The coherent coupling strength $g = \gamma /2$, in units of the dissipative coupling strength $\gamma$, satisfying the magnitude cascaded condition of Eq.~\eqref{eq:casc_cond_1}. We show increasingly strong dissipative coupling across the row, with $\gamma = \{ 1/2, 3/4, 9/10 \} \gamma_0$. Thick green lines: the completely reciprocal case, where no complex arguments are attached to the coupling constants [see the sketch above Fig.~\ref{cascadedpop}~(a)]. Medium pink lines: a chiral case, where $\theta_{12} - \phi_{12} = \pi/2$, leads to cascaded coupling between oscillators 1 and 2 [see the sketch above Fig.~\ref{cascadedpop}~(b)]. Thin cyan line: the completely cascaded case [cf. Eq.~\eqref{eq:casc_cond_2}], where $\theta_{nm} - \phi_{nm} = \pi/2$, leading to clockwise circulation [see the sketch above Fig.~\ref{cascadedpop}~(c)]. Lower rows: scaled local currents $J_{nm}/\gamma_0$, as a function of scaled time $\gamma_0 t$, for the three phases corresponding to the top row [cf. Eq.~\eqref{eqapp:current4}]. The solid, dashed and dotted lines represent $J_{12}$, $J_{23}$, and $J_{31}$ respectively. In the figure, the initial condition at $t=0$ is $n_1 = 1$, the gain into oscillator $1$ is $P_1 = \gamma_0/100$, and $P_2 = P_3  = 0$.  }
 \label{cascadedcurrent}
\end{figure*}

As the third application of the general theory, we consider cascaded configurations of the trimer. That is, where unidirectional or one-way coupling between some (or all) oscillators is possible~\cite{Gardiner1993, Carmichael1993}. We feed the system via the gain $P_1$ into the first oscillator, and allow for equal losses $\gamma_0$ across all oscillators, as is sketched in Fig.~\ref{cascadedsketch}. We include both coherent and dissipative coupling, which are of strengths $g$ and $\gamma$, and which are associated with complex arguments $\theta$ and $\phi$ respectively.

The interplay between the coherent and dissipative coupling can lead to the master equation of Eq.~\eqref{eq:master} mapping onto the celebrated master equation of cascaded quantum systems: that of one-way coupling between a source and target, with strictly no back action~\cite{Gardiner1993, Carmichael1993}. The equivalence of master equations occurs when the following two conditions are fulfilled~\cite{Supporting, Metelmann2015, Downing2019, Downing2020}
\begin{subequations}
\label{eq:casc_cond}
\begin{alignat}{3}
 g =&~\frac{\gamma}{2}, \label{eq:casc_cond_1} \\
 \theta_{nm} - \phi_{nm} =&~\begin{cases}
    \frac{\pi}{2}, & \lcirclearrowright~\mathrm{circulation}, \\
    \frac{3\pi}{2}, & \rcirclearrowleft~\mathrm{circulation},
\end{cases} \label{eq:casc_cond_2}
\end{alignat}
\end{subequations}
namely, both the strength condition of Eq.~\eqref{eq:casc_cond_1}, and the phase condition of Eq.~\eqref{eq:casc_cond_2}, must hold.

We calculate the mean populations $n_m = \langle b_m^\dagger b_m \rangle$ of each oscillator $m$ in this setup using both the transient and steady state solutions of Eq.~\eqref{eqapp:of_motion}. We plot the population dynamics in Fig.~\ref{cascadedpop}, where we set $g = \gamma/2$ throughout in order to satisfy the strength condition of Eq.~\eqref{eq:casc_cond_1}, and where the system is fed by the gain $P_1 = \gamma_0/100$. We show increasingly strong dissipative coupling $\gamma$ upon descending the rows, with $\gamma = \{ 1/2, 3/4, 9/10 \} \gamma_0$. The first column of Fig.~\ref{cascadedpop} presents a completely reciprocal case, where no complex arguments are attached to the coupling constants ($\theta_{nm} - \phi_{nm} = 0$), as is implied by the sketch above panel~(a). The population dynamics of the second (orange lines) and third (cyan lines) oscillators in panels (a), (d), (g) is therefore equivalent and there is no population imbalance. In the second column of Fig.~\ref{cascadedpop}, the configuration is such that $\theta_{12} - \phi_{12} = \pi/2$ and the other relative phases are zero, giving rise to unidirectional coupling between oscillators $1$ and $2$ [cf. Eq.~\eqref{eq:casc_cond_2}], as is highlighted by the red arrows in the sketch above panel~(b). The effect on the population dynamics is drastic, leading to significant excitation transfer from the first oscillator to the second oscillator (green $\to$ orange lines), which increases with increasing dissipative couping in going from panel (b) to (e) and (h). The final column of Fig.~\ref{cascadedpop} represents the most interesting case, with unidirectional coupling $\theta_{nm} - \phi_{nm} = \pi/2$ throughout the trimer, as is suggested by the sketch above panel~(c). Here the upper phase configuration of Eq.~\eqref{eq:casc_cond_2} is satisfied throughout the system, leading to a clockwise circulation of population $1 \to 2 \to 3$ (or, green $\to$ orange $\to$ cyan). This fact is most noticeable with strongest dissipative coupling in panel (i), where the population cycles are remarkably maintained, with minimal decay over time.

The associated global current $J$ is plotted in the top row of Fig.~\ref{cascadedcurrent}, using Eq.~\eqref{eqapp:current4}. The coherent coupling strength remains $g = \gamma /2$, satisfying the magnitude cascaded condition of Eq.~\eqref{eq:casc_cond_1}. We show increasingly strong dissipative coupling upon descending the column, with $\gamma = \{ 1/2, 3/4, 9/10 \} \gamma_0$. The thick green lines denote the completely reciprocal cases, where no complex arguments are attached to the coupling constants [see the sketch above Fig.~\ref{cascadedpop}~(a)], leading to zero global current in all top panels (a, b, c). The medium pink lines describe a chiral case, where $\theta_{12} - \phi_{12} = \pi/2$ and the other phase differences are zero, which leads to cascaded coupling between oscillators $1$ and $2$ [see the sketch above Fig.~\ref{cascadedpop}~(b)]. The thin cyan lines showcase the completely cascaded cases [cf. Eq.~\eqref{eq:casc_cond_2}], where $\theta_{nm} - \phi_{nm} = \pi/2$, leading to clockwise circulation [see the sketch above Fig.~\ref{cascadedpop}~(c)]. Clearly, the magnitude of the global current $J$ is higher with stronger dissipative coupling, best exemplified by Fig.~\ref{cascadedcurrent}~(c), which also showcases the formation of a non-negligible steady state current at long time scales $\gamma_0 t \gg 10$. Most notably, the global current $J$ is not sign changing in the top row of Fig.~\ref{cascadedcurrent}, despite the local currents $\{ J_{12}, J_{23}, J_{31} \}$ alternating in sign, as displayed in the lower rows of Fig.~\ref{cascadedcurrent}. This is because the local currents $J_{nm}$ are heavily weighted towards clockwise circulation for the nontrivial phase cases (pink and cyan lines) leading to a net positive global current.


\section*{Conclusions}
\label{Conc}

We have introduced a minimal open quantum systems model for directional circulation in a trimer of oscillators. Our general theory encompasses several special cases of interest, including a regime governed by a $\mathcal{PT}$ symmetric Hamiltonian, which displays nontrivial phases and novel population dynamics. We also reveal an arrangement allowing for completely unidirectional coupling throughout the system via a link to cascaded quantum systems, which displays the strongest signatures of chiral currents. This unidirectional coupling regime reveals the potentially beneficial influence of collective dissipation for chiral transport.

Our simple theory provides a foundation for more complicated simulations of chiral transport at the nanoscale, and highlights the possibilities for the experimental detection of chiral currents in a plethora of modern metamaterials, based for example upon photonic~\cite{Roushan2017, Roushan2017b, Owens2018, Ma2019} or plasmonic~\cite{Zohar2014, Lu2015, Barrow2016, Downing2017} excitations. The deployment of circuit QED platforms, which are especially amenable to artificial gauge fields~\cite{Peropadre2013, Baust2015}, is another appealing implementation, as are clusters of ultracold~\cite{Aidelsburger2011, Aidelsburger2013, Atala2014} and Rydberg atoms~\cite{Barredo2015, Lienhard2020}. One may envisage practical applications of our found chiral effects in a range of nonreciprocal nanophotonic devices, including isolators~\cite{Jalas2013, Sollner2015, Sayrin2015, Zhang2018}, circulators~\cite{Scheucher2016, Barzanjeh2017, Shen2018, Ruesink2018} and waveguides~\cite{Roy2017, Bello2019, Sanchez2019}.



\section*{Acknowledgments}
CAD acknowledges support from the Juan de la Cierva program (MINECO, Spain).
LMM and DZ were supported by the Spanish MINECO (Contract No. MAT2017-88358-C3-I-R) and by the the Arag\'{o}n government through the project Q-MAD. DZ also thanks the Fundaci\'{o}n BBVA, and the EU-QUANTERA Project SUMO. We are grateful to J.~C.~L\'{o}pez~Carre\~{n}o, A.~I.~Fern\'{a}ndez-Dom\'{i}nguez, and E.~del Valle for fruitful discussions.


\begin{thebibliography}{100}
\bibitem{mynoto}
This result follows since the Hamiltonian $H''$, as defined in Eq.~\eqref{eq:Ham_PT2}, is invariant under the pair of transformations $\mathrm{i} \to -\mathrm{i}$ and $2 \rightleftarrows 3$.
\end{thebibliography}
\end{document}